\documentclass[letterpaper, 10 pt, conference]{ieeeconf}  

\IEEEoverridecommandlockouts                              

\usepackage{stfloats}
\usepackage{float} 

\usepackage{booktabs} 
\usepackage{tabularx} 
\title{\LARGE \bf
In-Home Social Robots Design for Cognitive Stimulation Therapy in Dementia Care
}

\author{Emmanuel Akinrintoyo and Nicole Salomons 
\thanks{Emmanuel Akinrintoyo and Nicole Salomons are with Imperial College London
        {\tt\small , e.akinrintoyo23@imperial.ac.uk, n.salomons@imperial.ac.uk}}%
        }%

\usepackage{graphicx}
\usepackage{dirtytalk}
\usepackage{booktabs} 
\usepackage{amsmath}

\begin{document}

\maketitle
\thispagestyle{empty}
\pagestyle{empty}

\begin{abstract}
Individual cognitive stimulation therapy (iCST) is a non-pharmacological intervention for improving the cognition and quality of life of persons with dementia (PwDs); however, its effectiveness is limited by low adherence to delivery by their family members. In this work, we present the user-centered design and evaluation of a novel socially assistive robotic system to provide iCST therapy to PwDs in their homes for long-term use. We consulted with 16 dementia caregivers and professionals. Through these consultations, we gathered design guidelines and developed the prototype. The prototype was validated by testing it with three dementia professionals and five PwDs. The evaluation revealed PwDs enjoyed using the system and are willing to adopt its use over the long term. One shortcoming was the system's speech-to-text capabilities, where it frequently failed to understand the PwDs.
\end{abstract}

\section{INTRODUCTION}

Dementia is a syndrome that leads to a decline in cognitive functions such that individuals lose synchronisation with time, place, and person~\cite{greene1983reality}. This loss of synchronisation, known as reality disorientation~\cite{greene1983reality}, reduces the well-being and quality of life of PwDs, limiting their ability to live independently. The effects of disorientation become more severe as the syndrome progresses~\cite{gottesman2019behavioral}. 

Evidence-based non-pharmacological interventions, such as individual cognitive stimulation therapy (iCST)~\cite{orrell2017impact, ali2018individual} can alleviate the symptoms of dementia. iCST is a home-based individualised therapy for person-centered delivery tailored to the specific needs, interests, and preferences of a patient~\cite{orrell2017impact}. It consists of up to three 45-minute sessions weekly designed to be delivered by family carers in-home~\cite{ali2018individual}. This involves reminiscing about past memories and discussing daily living activities and current news. iCST has been shown to improve cognition and the quality of life of PwDs~\cite{orgeta2015individual}.

Despite iCST's potential, it has some limitations. A randomised control study showed that, despite being effective, many participants did not complete the recommended number of weekly sessions~\cite{orrell2017impact}. The main reason for non-adherence was the overload of caregiver responsibility~\cite{ali2018individual}, where family caregivers needed to combine its delivery three times a week with other care responsibilities. An additional difficulty was the need for specialised training to deliver the structured iCST program.

\begin{figure}
    \centering
    \includegraphics[width=0.4\textwidth]{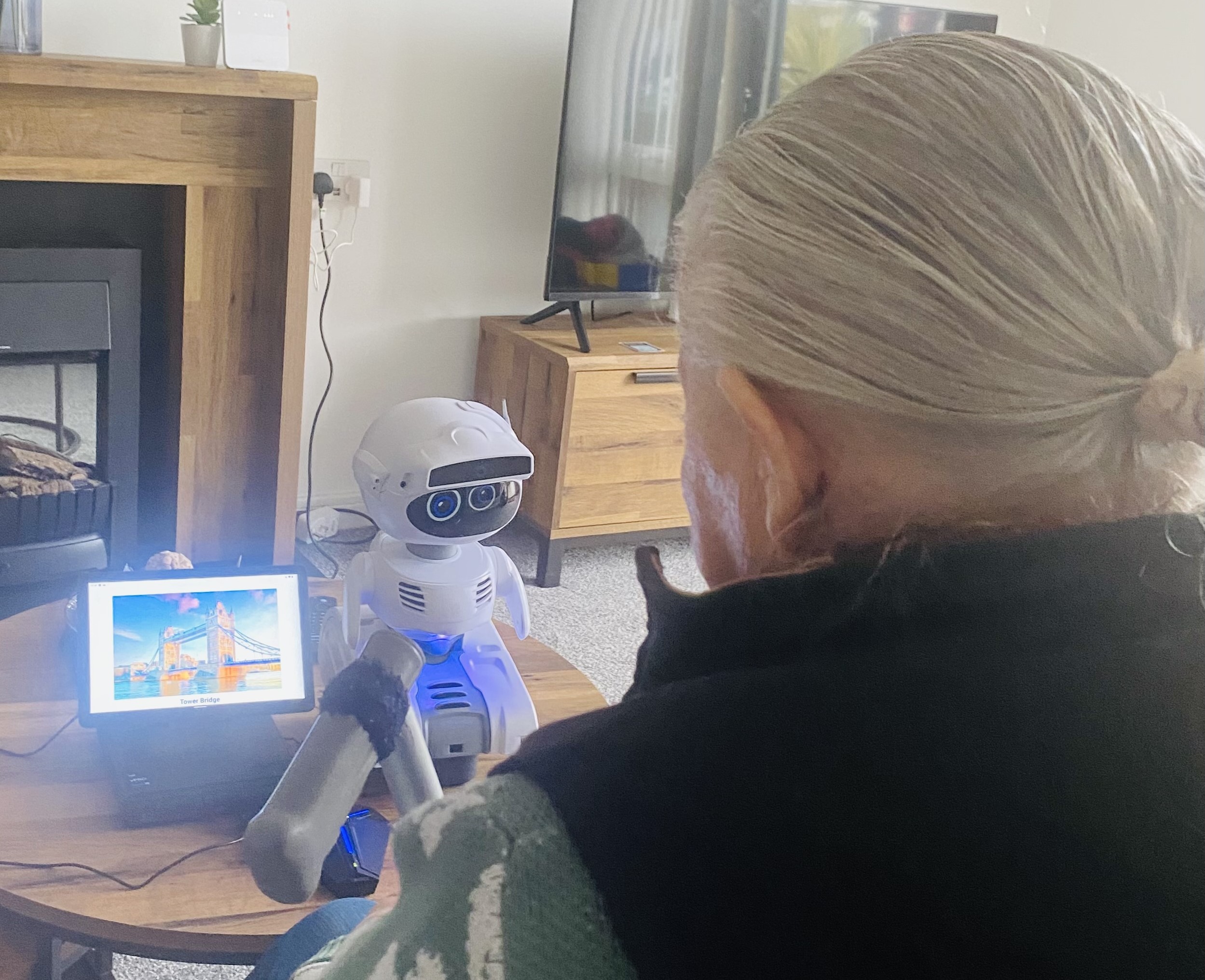}
    \caption{We designed a robotic system that provides individualised cognitive stimulation therapy to people with dementia in their own homes.}
    \label{fig:square_image}
\end{figure}


In this work, we developed a robotics-based system for iCST delivery. It can deliver daily cognitive activities on themes based on the preferences and interests of a PwD. Likewise, it provides consistency and accessibility by being stationed at home 24/7. The system demonstrates the significant potential of a social robot to increase iCST adherence. This paper presents the iterative user-centered design process with relevant stakeholders (PwDs, their families and dementia care professionals) for the design of an in-home social robot for iCST delivery. This process first involves a requirement gathering (Phase 1) in which the design needs, preferences, and expectations of 16 dementia caregivers and professionals are gathered. Key design goals include personalisation, autonomy, and a focus on creating personal connections with PwDs rather than correcting their mistakes. 

Subsequently, a prototype design of the system was developed (Phase 2), consisting of a robotic system that provided six different iCST activities. The prototype was brought to several independent living facilities (Fig. \ref{fig:square_image}) and tested in one-to-one sessions by three dementia professionals (Phase 3) and five PwDs (Phase 4). These sessions revealed that users found the system easy to use and beneficial for cognitive stimulation. Lastly, we provide insights and lessons learned for those who plan to conduct research on robotic therapeutic systems for PwDs.


\section{BACKGROUND}
This section provides a background on cognitive stimulation therapy and robot-assisted therapy for dementia.

\subsection{Individual Cognitive Stimulation Therapy (iCST)}
CST is an evidence-based psychological intervention for improving the cognitive functions, well-being and quality of life of individuals with mild to moderate dementia~\cite{morley2014cognitive}. Conventionally, CST is delivered in group settings for PwDs, unlike iCST, which is its personalised, one-on-one counterpart~\cite{orrell2017impact, ali2018individual}. iCST is delivered by informal carers (that is, the family members). The sessions involve discussing various themes such as popular faces, places, food, and household items~\cite{yates2016field}. The activities are aided by visuals such as images, text and puzzles. The benefits of iCST include time flexibility, as it can be delivered at home at any time without the need to travel to a specific location. Notably, the strength of iCST lies in its person-centric approach~\cite{rai2021field}. This ensures that the activities delivered cater to the specific cognitive needs and interests of a person living with dementia.

The randomised control study that sought to investigate the efficacy of iCST revealed that its one-to-one delivery approach introduces a constraint~\cite{orrell2017impact}. The individualised approach can be emotionally draining for caregivers, especially considering the time commitment needed for regular delivery. iCST requires sustained focus, energy and engagement to be delivered effectively~\cite{moniz1998preliminary}. Furthermore, emotional stress from caregiving, especially during cognitive decline in loved ones, can make it difficult for caregivers to look after their own well-being. This could lead to caregiver burnout, thus impacting the quality of care they provide and their adherence to the intervention's delivery.

\subsection{Robotics for Cognitive Stimulation Therapy (CST)}

Robot-aided therapy studies for PwDs have primarily focused on CST robotic delivery involving groups of patients~\cite{castellano2022detecting, cruz2020social, shukla2017effectiveness}. Castellano et al.~\cite{castellano2022detecting} had a group of eight participants with mild cognitive impairment in a three-week program (for 35 minutes weekly in a care facility). The study focused on detecting the emotions of PwDs. It revealed that participants displayed positive emotions during sessions and emphasised the need for personalised CST. Cruz-Sandoval et al.~\cite{cruz2020social} also involved eight PwDs in their eight-week study. It found social robots effective at decreasing dementia-related behavioural changes (e.g. agitation).

Other studies, such as Khan~\cite{khan2023humanoid}, were conducted in a lab setting without the target population~\cite{chan2010promoting, cobo2021development, minhaj2023development}. It found that robotics-based CST improves cognition and reduces behavioural symptoms. Chan et al.~\cite{chan2010promoting} developed a robot to deliver cognitive games for PwDs. They tested it in a lab with healthy adults and found it effective in engaging the participants. Minhaj et al.~\cite{minhaj2023development} developed a brain-computer interface (BCI) feedback system to assess the patient’s mental activity. The system was tested in an online setting with three young people. 

The producer-consumer style approach creates a gap in technology adoption, as the core needs of the PwDs may not be captured in the design. Additionally, the designs fail to capture how PwDs would interact with them at home or care centers~\cite{chan2010promoting, khan2023humanoid}. The dynamics of an in-home deployment differ from controlled settings such as a lab. This may make the designs unsuitable for in-home deployment or long-term use, especially when they are tested without the target population. Hence, this research adopts a user-centered approach to develop the system for long-term, in-home use.

\section{Phase 1: Interviews with dementia caregivers and professionals (DCPs)} 

Consultations were held with DCPs to gain insights into dementia care, as they are familiar with the challenges and symptoms of PwDs. PwDs were excluded from Phase 1 consultations to avoid potential difficulties in understanding the social robot concept and to reduce the stress of participating in multiple interviews.

A group consultation was conducted with three DCPs at an independent living facility, where we identified a key research gap in dementia care that could be addressed by a social robot. The DCPs noted that most PwDs spend much of their time alone. PwDs receive one-hour daily visits from DCPs who help with daily living tasks, but for the rest of the day, many PwDs have minimal engagement. This lack of stimulation contributes to rapid cognitive decline, particularly for those living alone. Although some PwDs use devices such as smart speakers and dementia memory clocks for stimulation, the effects are short-lived. Hence, DCPs emphasised the need for a long-term solution to provide consistent daily stimulation.


An initial consultation was held with a lead CST researcher to inform the development of the study. Subsequently, 14 semi-structured interviews were held with stakeholders. This included formal and informal caregivers, therapists, and psychologists, who work with PwDs daily. The interviews included questions on their past experiences with disorientation, cognitive stimulation activities and therapy, and the potential of using a social robot to provide cognitive stimulation. The interviews were conducted via Microsoft Teams. Table \ref{tableall} presents the details of the 14 distinct persons interviewed based on their roles and experience in dementia care. Each participant was interviewed for one hour and received compensation of 15 pounds for their time. The study had ethical approval (ethics no: 6871649).

\begin{table}[t]
\centering
\caption{Participant roles, gender, experience (yrs), and IDs.}
\label{tab:participant_info}
\footnotesize
\renewcommand{\arraystretch}{1.05}
\begin{tabularx}{\linewidth}{p{3.2cm} p{1.2cm} p{1.8cm} X}
\toprule
Role & Gender & Exp. & IDs \\
\midrule
Formal caregivers & 3F & 4.5–15 & FC1–3 \\
Informal caregivers & 4F & 10–21 & IC1–4 \\
Occupational therapist & 1F & 20 & OT1 \\
Dem. nurse practitioners & 2F & 10–24 & DNP1–2 \\
Well-being \& tech leads & 1F, 1M & 11–15 & WTL1–2 \\
Mental health practitioners & 1F, 1M & 6.5–24 & MHP1–2 \\
\bottomrule
\label{tableall}
\end{tabularx}
\end{table}

\vspace{-0.25cm}

\subsection{Interview Findings}
Content analysis was used for all the interviews conducted in this study.

\subsubsection{Human Delivered CST}\label{humans}

The majority of stakeholders had experience in delivering CST. Five had delivered group CST only, four had experience delivering both group and individual CST, and two had delivered iCST only. The remainder (three) had no practical experience in delivering CST but had some knowledge of it based on its delivery at their dementia care facility or having taken a loved one to CST sessions.

All participants found CST effective in stimulating cognitive function, enhancing orientation, and revitalising engagement, based on their experiences with clients and loved ones. WTL2 said that \textit{\say{CST can be quite magical}} and \textit{\say{it has had a massive impact on the people}}. It enabled some PwDs to be \textit{\say{switched back on}} and \textit{\say{gives them some quality time with their family member}} (WTL2). They became \textit{\say{so much better orientated to time, place, and person, even those with quite a much more progressed version of dementia}} (FC3). 

They also noted the challenges with its delivery such as cost -- \textit{\say{it does cost a lot of money for the carers to deliver the sessions}} (IC4), lack of time -- \textit{\say{it just wouldn't be feasible for me to go into someone's home three times a week for half an hour [of iCST]}} (OT1), and staff confidence -- \textit{\say{when you hear cognitive stimulation therapy, the member of staff thinks, oh my goodness, this sounds so technical}} (WTL2). IC1 emphasised that caring for a PwD means that \textit{\say{you don't get any \say{me time}}}, leading to mental challenges when delivering iCST. In addition, \textit{\say{family members have expectations for the person that they expect them to be able to do this word recall [activity]}} (DNP1). 

All the challenges noted were grouped into four categories, with the number of participants who stated them. They include (i) staffing issues such as lack of staff time, interest and confidence (11), (ii) resource issues such as cost of resources and accessibility (4), (iii) program structure and consistency such as lack of structure in delivery (3), and (iv) emotional and mental stress of informal caregivers (2).

\subsubsection{Social Robots for iCST}\label{finding1}\hfill

\noindent\textbf{Potential:} All the participants agreed that technology is a potential solution for human-delivered CST delivery, with MHP1 stating that \textit{\say{we have to accept technology is good, it's a way to move forward}}. All the interviewees foresee a future where robots will be part of the care intervention plan for PwDs to provide regular iCST activities. They considered it as a solution that \textit{\say{can really make a difference}} and \textit{\say{is almost inevitable, particularly with the advent of AI [artificial intelligence]}} (MHP2). This is because robot-delivered iCST can \textit{\say{definitely free up staff time}} (FC2) and provide \textit{\say{consistency that humans can't}} (WTL2). Furthermore, a robot, compared to a person, could give less judgemental responses because the robot \textit{\say{is not going to get tired, it's not going to get bored or fed up with the person, it's not going to roll its eyes because the person with dementia is repeating themselves}} (WTL2). However, we acknowledge potential bias in the robot's responses.

\noindent\textbf{Personalisation:} All participants identified personalisation as a vital design feature. One person stressed that \textit{\say{because you want to know the person better, without that [personalisation] it's not iCST}} (MHP1). Likewise, it will ensure that \textit{\say{they are likely to stay engaged}} (FC3). 

The most commonly cited methods by participants for the robot to personalise its utterances and behaviours were classified into five categories. The categories include interests and hobbies (8), personal information (5), personal history (4), skills and abilities (1) and personal traits (1). Interests and hobbies include likes, dislikes, music, pets, and social activities. Personal information includes name, literacy level, nationality, and family. Personal history includes past occupation, childhood, education and medical history. 

\noindent\textbf{Audio recording:} When asked about audio recordings of iCST sessions, 11 interviewees accepted their use. This included WTL2, who described that \textit{\say{it would be good if the robot could have the functionality but the person could opt in or out}}. Three believed it was dependent on the user's preference because \textit{\say{some people don't want to be recorded at all but others are okay with audio}} (DNP1). While audio recording might be acceptable, WTL1 noted that videos are an exception because \textit{\say{family members want cameras removed from [independent living] apartments, they don't agree to it}}.


\noindent\textbf{Robots vs caregivers:}
When asked if PwDs would complete more iCST sessions if it were being delivered by a robot compared to a caregiver, 13 participants suggested that a robot would enable PwDs to engage in more regular cognitive stimulation activities. FC2 described that \textit{\say{it will definitely free up staff time}} because \textit{\say{it is not taking time out of anybody's day}} (WTL1). However, DNP1 felt PwDs may complete fewer sessions because the \textit{\say{emotional connection [with humans] will be missing}}, thus potentially leading to lesser engagement. Nevertheless, they cited that \textit{\say{it's not possible to have a person go into lots of people's homes to deliver one-to-one CST}}.

\subsubsection{iCST Session Design} \label{finding2}\hfill\\
\noindent\textbf{Duration:} 
When asked whether they would prefer 30-minute or 60-minute iCST sessions, a duration of 30 minutes was preferred by 12 interviewees. This is because \textit{\say{an hour is too long and two hours is absolutely too long}} (MHP2). However, two participants preferred to have longer sessions that span up to one hour, including IC2, who discussed that \textit{\say{she [PwD] can keep going for quite a few hours}}.

\noindent\textbf{Number of sessions:} 
Participants were asked how many daily sessions a robot should deliver. Twelve interviewees suggested two cognitive stimulation activities daily. IC4 stated that it is necessary because \textit{\say{a lot of people with dementia are left by themselves for long periods of time with no stimuli at all}}. Two participants opted for one session per day design \textit{\say{to avoid it being too overwhelming}} (FC3).

\noindent\textbf{Time of day:} 
Interviewees were asked what times of the day iCST should be delivered. Ten participants preferred the morning, with eight participants preferring late morning (10-12\,\text{am}) \textit{\say{so it's not too early for people who don't get up early}} (FC3). Eight participants suggested afternoon times, with four of them preferring sometime around 2\,\text{pm}. Four participants believed it is person-dependent \textit{\say{because everyone is different}} (MHP1) and \textit{\say{some people are more lucid at certain times of the day}} (WTL1). WTL2 suggested PwDs \textit{\say{can have a session whenever they need some stimulation}}.

\noindent\textbf{Multi-person design:}
Participants were asked if they would prefer a design where PwDs complete a session independently or one that allows loved ones, or caregivers to join in when they are available. Eleven interviewees envisioned a design whereby they could join when they wanted to. This is \textit{\say{because it makes it quite social}} (WTL2), and it will enable the \textit{\say{person with dementia to trust the robot}} (MHP1). All informal caregivers preferred this.
Others (three) opted against it, preferring to watch or supervise the sessions.

\noindent\textbf{Feedback:} 
All the interviewees suggested that feedback on how the session went should be sought from PwDs on their daily iCST sessions. Four participants suggested that if two sessions were to be delivered daily, then feedback should be sought only for the second session. Others (10) thought it was important to ask for feedback every session.

\noindent\textbf{Stand vs. sit:} Ten participants preferred to have a user sit while completing a session as opposed to standing for \textit{\say{the person to be more comfortable}} (DNP1). IC3 preferred standing. 
Three interviewees suggested both options with \textit{\say{one activity standing, one activity sitting}} (WTL1). 

\section{Design Goals}

Based on the interviews, seven design goals were identified. The first three are iCST principles~\cite{yates2015development, yates2016development}, while others are derived from the interview findings.

\noindent\textbf{I. Connect, not correct: } 
The robot should focus on developing a connection with the PwD throughout the activities rather than on correcting their mistakes. This also addresses the concerns of some stakeholders who felt that the emotional connection present with humans will be missing. 

\noindent\textbf{II. Stimulation, not training:} To aid PwDs in completing more sessions, the activities should focus on stimulation without seeking to train specific aspects of cognition. Stimulation would make them \textit{\say{better orientated to time, place, person}} (WTL2).

\noindent\textbf{III. Opinions, not facts:}
The activities are designed to obtain users' opinions on various topics, places, and objects. Hence, the emphasis is on opinion-sharing and discussion, which makes them suitable even for users who may forget information quickly, as IC1 described, \textit{\say{you tell her something then when you've finished the sentence she's forgotten}}.



\noindent\textbf{IV. Autonomy:} \label{autonomy}
An autonomous design alleviates the responsibility of the caregivers \textit{\say{to free up staff time}} (WTL1). Many interviewees noticed the potential of robots to free up staff time; therefore, the robot must operate autonomously to prevent staff from spending time operating it instead. The system needs to operate in PwDs' homes and must power itself on or off and deliver the sessions independently, especially since many older people do not know how to operate robots or technology. Caregivers or PwDs should not have to manage any system operations. 

\noindent\textbf{V. Privacy:} 
Following the interviews, it was determined that data privacy is a crucial design aspect because audio recordings of the interactions will be collected. This will be achieved using a private cloud to ensure that users' data will remain locally within their homes without using a third-party cloud service for storage. Video data will not be collected.

\noindent\textbf{VI. Personalisation:} 
The system will adopt a personalised design approach using the recommendations of the stakeholders. The robot will better engage PwDs by tailoring the interactions to their unique preferences, history, and interests. This will significantly enhance the effectiveness and comfort of care for PwDs.

\noindent\textbf{VII. Multi-person design:} 
The design must accommodate multiple participants, such as PwDs and their loved ones, during CST sessions. This will strengthen their social connections and make the interactions more meaningful. It will also eliminate biases or apprehensions that PwDs may have about participating alone due to their cognitive challenges.

\section{PHASE 2: Prototype Design} 

The prototype of the system and the design of the iCST sessions were based on the interviews and the design goals. 

\subsection{iCST Activities}
iCST activities were developed based on its recommended manual~\cite{icst_pack}. Each activity contains reminiscence elements with some discussion. The activities developed include (i) popular faces, (ii) common sayings, (iii) famous places, (iv) reminiscence/childhood, (v) categorising objects, and (vi) orientation. The system evaluation in sections \ref{Phase3} and \ref{Phase4} was done with the famous places and common sayings activities. 

The famous places session evokes memories of visits to popular and historic buildings. A PwD is shown an image of a popular site, such as Buckingham Palace and asked whether they can identify it. The common sayings activity involves asking users to complete proverbs or popular quotes. Examples include completing the saying \textit{\say{a picture is worth a thousand \underline{words}}}. Follow-up questions are asked based on each image or quote for an engaging discussion. 



\subsection{iCST Sessions}

The robot’s main interaction with the user is the delivery of iCST sessions. The robot can deliver the pre-programmed cognitive sessions twice daily, with each session lasting 30 minutes. Based on the interview findings, the system is set up for users to stay seated during the interaction. A user can choose to start an activity when they decide to, or an activity can commence at a specified time. Furthermore, the conversations are recorded for analysis, subject to a user's consent, with a log of which session was completed.

At the start of a session, the robot greets the user and engages them by playing one of their favourite songs or discussing a current news topic to capture their attention. This sets the mood of the activity and allows the user to relax. The robot then provides basic orientation information about the time, date and place (city and country). Subsequently, an activity is introduced along with its description and what is expected from the user. To conclude, feedback and opinions on the session are sought to gauge the user's interest and enjoyment before thanking them. 


\subsection{Hardware and Software}

\begin{figure}
    \centering
    \includegraphics[width=0.37\textwidth]{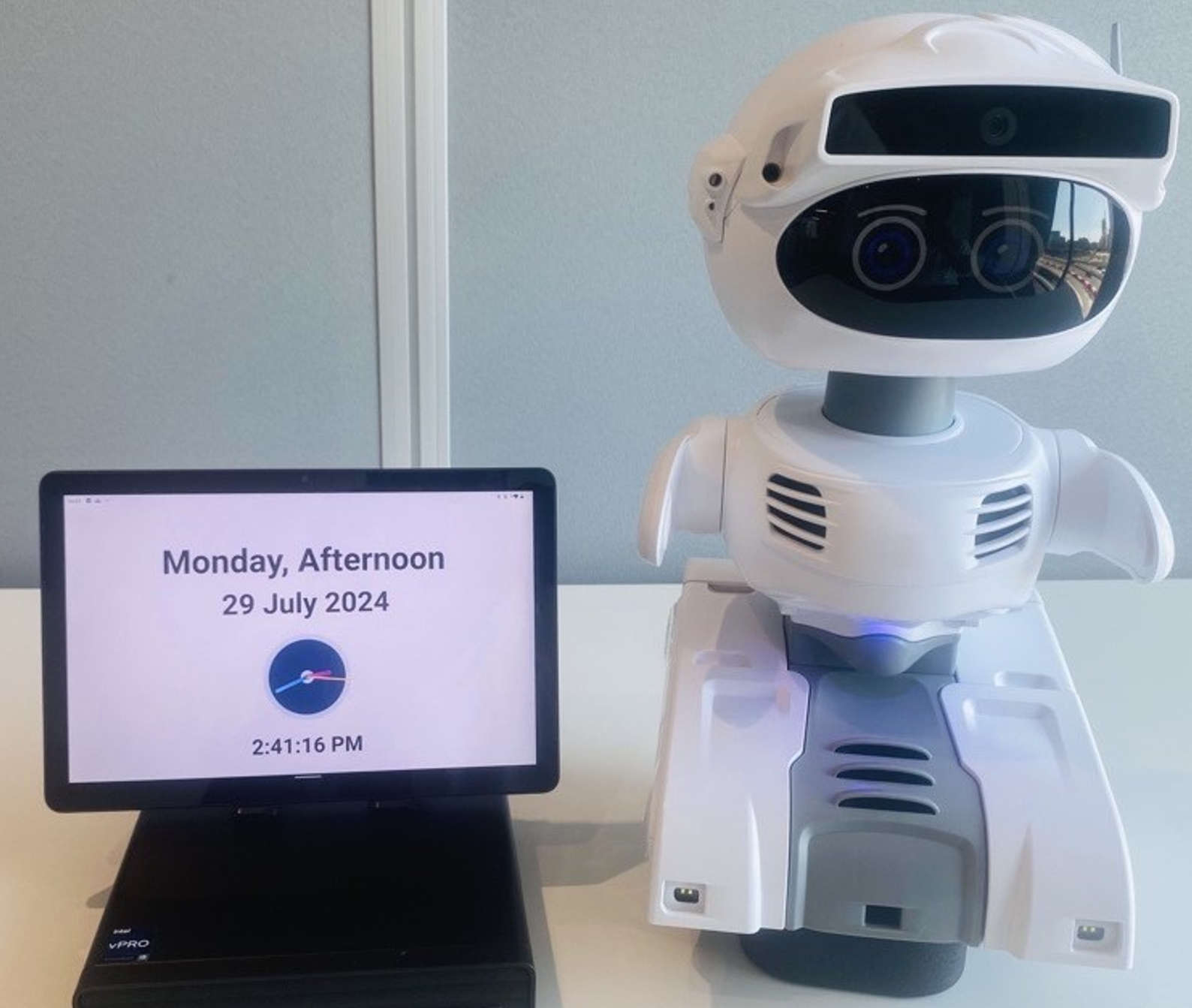}
    \caption{The system setup consists of a Misty robot, a table and a mini-computer for delivering iCST sessions at home.}
    \label{fig:misty}
\end{figure}

The system consists of a mini computer, a tablet, a microphone, and a Misty robot as portrayed. The computer controls the tablet and the robot. The tablet provides a means to display images and texts relevant to each activity. None of the activities requires the user to operate the tablet. Instead, the user communicates verbally with the robot. An extra microphone is used to capture audio responses from the user. The Misty robot~\cite{Misty_Robotics_2023} was chosen because of its suitability for human interactions and low cost compared to alternative social robots. Misty is a gender-neutral table-top-sized robot of approximately 14 inches in height and 6 pounds in weight. 

Misty's speech was generated using OpenAI's text-to-speech functionalities and Alloy's voice for a more natural sound. Google's web speech API is utilised for the speech-to-text (STT) functionality. The user’s responses are analysed such that the nouns and pronouns are extracted from the input sentences. This is then checked with a set of predefined correct answers. NLP (natural language processing) allows for wrong pronunciations by the user or close transcriptions of the speech API (application programming interface) with a similarity index of 70\%.

\section{Phase 3: System Testing with Dementia Professionals}\label{Phase3}
Three professionals (2 WTLs and 1 formal caregiver) completed 20-minute one-to-one sessions with the robot. During each session, participants completed seven famous place activity questions, along with one follow-up question, and four common sayings activities, each with two follow-up questions. A 10-minute interview followed this to assess their experience and gather recommendations for the system.

\begin{table}
  \caption{PwDs' profile including their gender, age, ethnicity, whether they have prior CST experience and their level of technological familiarity.}
  \label{tab:pwdTab}
  \begin{tabular}{cccccp{1cm}} 
    \toprule
    IDs & Gender & Age & Ethnicity & Prior CST & Tech.\\
    \midrule
    PwD1 & M & 84 & Afro-Caribbean & Yes & Minimal \\
    PwD2 & M & 87 & Afro-Caribbean  & No & Minimal  \\
    PwD3 & F & 80 & White European  & No   & Minimal      \\
    PwD4 & F & 76 & White European  & Yes   & Moderate       \\
    PwD5 & F & 89 & White European  & No    & None         \\
    \bottomrule
  \end{tabular}
\end{table}

\noindent\textbf{Usability:} 
Using ISO/IEC 9241-11 standards, participants completed tasks successfully and on time, showing effectiveness and efficiency, and reported clear instructions and ease of use, indicating satisfaction. One noted the \textit{\say{words on the tablet might be an issue [for PwDs in later stages]}}.
\newline\noindent\textbf{Perceived effectiveness:}
The participants found the sessions stimulating. One described that it \textit{\say{helped to think about trips that I have been to}}. The participants also discussed that the robot was helpful in completing the tasks \textit{\say{especially when I answered the question wrong}}.
\newline\noindent\textbf{Preference and comparison:}
One of the professionals found the experience comparable to human-delivered CST, noting that \textit{\say{it will be better for one-to-one [sessions than human carers], but I don’t think we can replace humans}}. Another discussed that \textit{\say{I’ve seen other examples aimed at more advanced dementia and it felt condescending, whereas this is more fun and makes you relaxed}}.
\newline\noindent\textbf{Comfort and engagement:}
All the participants felt comfortable during the session with the robot and enjoyed the activities. The professionals stated that they felt more relaxed after the session. They believed the iCST sessions helped to improve their mood as it was enjoyable for them to do.
\newline\noindent\textbf{Suggested Improvements:}  One interviewee stated the PwDs \textit{\say{may struggle with the level of the voice because they might speak low to the robot}} and they may not \textit{\say{find the mic interesting [to use]}}.
Lastly, one of them mentioned that Misty should \textit{\say{acknowledge the answer to have more of a flow in the conversation}} with a faster response time. Another participant reiterated this too before noting that \textit{\say{with dementia you wouldn’t want it to be faster}}.

\section{Phase 4: System Testing with PwDs}\label{Phase4}

Ethical approval was obtained for the study (ethics no: 7091501). One-hour in-person sessions were conducted at four independent living facilities. Table \ref{tab:pwdTab} presents the profile of the five PwDs who participated. They had mild to moderate dementia with sufficient capacity to consent for themselves. Each session was split into two: (i) a 20-minute one-to-one session completing the famous places and common sayings iCST sessions, as described in the previous section, and (ii) a 40-minute interview. Each participant received 15 pounds. 

\subsection{Statistics of Sessions}

Table \ref{tab:Evalresuls} presents the results of the sessions with the robot, showing the number of actual correct answers the user got (from 11 questions), the proportion of questions Misty marked as correct, and the number of times Misty asked the user to repeat themselves (27 total responses, including 15 follow-up questions and a request to know the user's name).



The STT worked better for the DCPs compared to the PwDs, indicating that the system could understand the speech of healthy persons much better. The STT failed mostly during follow-up questions. This often occurred because of the PwDs' speaking style. They often had a low voice response and mumbled their words. Sometimes, they took pauses while speaking, struggled to process their speech correctly or took too long to think about the correct response to a question. The system detected these moments of silence as the end of their responses. Also, for a landmark (the Palace of Westminster), there was a difference between the official name and the general public name (Parliament), which also led to the system incorrectly classifying the answers.

\begin{table}
  \caption{System evaluation of one-to-one iCST sessions with an assessment of how many correct answers each user gave, those the robot considered as accurate, and the STT's performance.}
  \label{tab:Evalresuls}
  \begin{tabular}{cccc} 
    \toprule
    IDs & Correct(\%) & Marked correct(\%) & STT Fail (\%) \\
    \midrule
    DCP1 & 63.6 & 54.5 & 14.8   \\ 
    DCP2 & 100 & 81.8 & 3.7  \\ 
    DCP3 & 54.6 & 36.4 & 37   \\  
    PwD1 & 27.3 & 18.2 & 51.9   \\  
    PwD2 & 0 & 0 & 100   \\
    PwD3 & 36.4 & 27.3 & 44.4        \\
    PwD4 & 45.5 & 36.4 & NA    \\ 
    PwD5 & 54.6 & 36.4 & 59.3     \\
    \bottomrule
  \end{tabular}
\end{table}

\subsection{System experiences}

\noindent\textbf{PwD1: }He claims to struggle with technology and, therefore, mostly ignores it in his day-to-day life. However, he was willing to try our system after a demonstration by the researcher because CST is \textit{\say{good for you}}. PwD1 found the instructions clear and easy to understand. He noted that the interaction sometimes helped him to stimulate his memory because it \textit{\say{helped identify the places you know}}. PwD1 struggled with speech difficulties, which made it difficult for the system to hear him well often. 

He wished the system were more personalised, as he wished to have more correct answers. 
Hence, he asked for popular places in his locality to be included because \textit{\say{I don’t remember them [other places]}}.

\noindent\textbf{PwD2: }PwD2 is an independent and mobile man with fewer speech capabilities. The system had difficulty processing his speech as he spoke quietly, and he could not answer any of the questions. He needed a lot of time to process the information and understand the content. The robot would often continue the conversation while he was trying to speak due to his low voice. This indicates that existing speech systems are not ideal for people with language difficulties. 

PwD2 found it more interesting to explain the proverbs than to describe the places. He had difficulty understanding some questions in the session. Thus, he constantly read the text to understand. However, he felt relaxed and engaged throughout, trying to answer the questions. He also found it helpful to have his formal caregiver around during the session. The caregiver gave him some reminders to speak louder when answering. This revealed that a multi-person design is ideal for persons with less speech and language. 

\noindent\textbf{PwD3: }PwD3’s session strongly highlighted the second design goal (stimulation, not training). She noted that \textit{\say{I can walk out of this door and not remember anything}}. She discussed that \textit{\say{I enjoyed it, but the thing about it is that, unlike other tests, nobody takes into account the person that has no memory}}.

The famous places activity had multiple images of places where she had worked. She stated that \textit{\say{I know all the buildings and I’ve worked in most of them, but now I can’t remember even though I know them}}. While she could not remember the buildings' names, it started conversations about her previous career. This surprised her formal caregiver, who said \textit{\say{they were things I didn’t know about her that those images showed, it did help her to think and talk about those places}}. PwD3 responded that \textit{\say{I have not talked for years about the aspects of my job we talked about}}.

\noindent\textbf{PwD4: }PwD4 was previously quite familiar with technology, having taught children computing --\textit{\say{I was in charge of computers}}. She could interact effectively with the system, with the correct voice pitch and tone for Misty to understand. PwD4 was fascinated with the system and noted that \textit{\say{I love this machine}}. The activities helped her to establish a connection with her previous career (in line with the first design goal to connect, not correct). 


\noindent\textbf{PwD5: }PwD5 is an older person who \textit{\say{would rather be active, not just sitting and doing nothing}}. While she had no previous experience with CST, she considered the session as \textit{\say{fun, it was good}}. She especially enjoyed the common sayings activity. In the famous places activity, she struggled with remembering the specific names of the places, and instead could describe these locations in detail. She struggled with her low voice level, which was not loud enough. This impacted how well the system could transcribe her answers and responses. Hence, Misty asked her to repeat responses frequently. Despite this, she did not find it annoying.

\subsection{Interview Findings}

Here we present the main interview findings. Due to a health challenge, PwD4 had to leave after the robotic CST session. Therefore, we do not have interview data for them.

\noindent\textbf{Personalisation:}
All the PwDs agreed with having the system personalise to their preferences. 

\noindent\textbf{Audio recording:}
All PwDs considered it necessary to record the sessions because \textit{\say{because you’ve [researcher] got to have summary}} (PwD4).

\noindent\textbf{Multi-person design:} 
PwDs, including PwD3, preferred to \textit{\say{do with others}} for a better chance of knowing the answers.

\noindent\textbf{Robots or caregivers:} Most participants would rather do it with the robot, mainly because \textit{\say{people [carers] here don’t have time. The only person who would do it is [my carer], and she's very busy}}. PwD5 preferred to do with a carer.  

\noindent\textbf{Duration:}  
All participants preferred a shorter session. PwD5 preferred 20 minutes while others chose 30 minutes, so \textit{\say{you don’t get bored or repeat things}} (PwD3). 

\noindent\textbf{Number of sessions:} 
One session a day was the most common option. PwD3 preferred twice daily while PwD4 noted that it would depend on \textit{\say{what I'm doing in the day}}. However, she emphasised that \textit{\say{people will rely on it more in the winter times and do more times}}.

\noindent\textbf{Time of day:}
PwD1 and PwD4 preferred 2pm\,\text{pm} \textit{\say{because that’s the time you have to relax. Sometimes in the morning, you have appointments}}. PwD3 preferred morning, while PwD5 did not have a specific preference. 

\noindent\textbf{Stand vs. sit:}
All participants preferred to sit \textit{\say{because standing up too long gives me a lot of pain}} (PwD1).

\section{Discussion}

\subsection{Insights}

\noindent\textbf{Personalisation} - 
While the system has personalisation capabilities, the iCST sessions do not include personalised content. Preferrable personalised content for future iterations includes images of a user's previous travel experiences, work history, and popular sites within their region, especially those in close proximity. When PwDs could not identify a specific location, they consistently referred to well-known local landmarks near their vicinity.

\noindent\textbf{Connect, not correct} - 
The robot created connections with each user. Moreover, it fostered the bond between PwDs and their carers. After the session, PwD4 continued a lengthy discussion with her carer on how robotics could help young people learn better. Instead of seeking to correct mistakes, it encouraged users when they missed a question. It responded with phrases such as \textit{\say{you have almost cracked it}} or \textit{\say{you are on the right path}}.


\noindent\textbf{Speech-to-Text} - 
Current STT APIs are not robust enough to understand the nuances of PwDs' speech. A crucial limitation of the system occurred when a PwD took a long pause when a question was asked or answered questions slowly. The system interpreted them as signals that the user had stopped talking. We consider the improvement of STT for dementia as a major open gap for future research. Furthermore, future systems should deliver more natural conversational interactions, such as additional personalised follow-up questions. 

\noindent\textbf{Entrainment and accommodation} - 
The participants adjusted their communication and speaking styles based on the robot's capabilities (validating the communication accommodation theory~\cite{giles2023communication}). They often modified how they enunciated or phrased some words to ensure Misty understood them better. Whenever it failed to hear them, they generally slowed down their speech. DCPs aligned more after a few questions, while PwDs defaulted to their normal speech style after some subsequent responses. Future work can implement machine learning models that adapt the robot's interaction style based on a user's real-time interaction behaviour. It will personalise the response speed of the robot while taking into account how often a PwD takes pauses while speaking.

\noindent\textbf{Open ended questions} - 
While PwDs often struggled to remember the places or sayings correctly, they often provided descriptions for them instead. Hence, the follow-up questions were profound for engaging with PwDs. An essential element could be to ask participants to describe the places or what they can see in the image. Future work will integrate questions from their memories of previous careers and hobbies. 

\subsection{Future Work}
We will create a second prototype with improved STT capabilities and personalised questions and responses for long-term deployment in PwDs' homes.

\bibliographystyle{IEEEtran}
\bibliography{IEEEabrv,bibliography}


\end{document}